# Translation-Rotation Coupling in the Dynamics of Linear Molecules in Water

**Anjali S Nair, Puja Banerjee, Sarmistha Sarkar and Biman Bagchi\***

Solid State and Structural Chemistry Unit, Indian Institute of Science,
Bangalore, India – 560012

Corresponding author's E-mail : profbiman@gmail.com ; bbagchi@iisc.ac.in

*Abstract*

*We study by computer simulations the coupled rotational and translational dynamics of three important linear diatomic molecules, namely, carbon monoxide (CO), nitric oxide (NO) and cyanide ion ($CN^-$) in water. Translational diffusion of these molecules is found to be strongly coupled to their own rotational dynamics which in turn are coupled to similar motions of surrounding water. We examined the validity of hydrodynamic predictions and found them to be largely insufficient, particularly for rotational diffusion. A mode coupling theory approach is developed and applied to understand the complexity of translation-rotation coupling.*

I. **INTRODUCTION**

While a large number of theoretical studies have been devoted to translational diffusion of spherical tagged molecules in solvents, for example, a solute sphere in binary Lennard-Jones mixture, studies of non-spherical molecules have drawn relatively less attention till this date, despite the fact that the over-whelming majority of solute-solvent systems consist of non-spherical molecules.[1–5] For example, methanol and ethanol in water represent the common system where molecules are non-spherical.[6] Chlorobenzene is another non-spherical molecule which is widely used in experimental studies on glass-transition. The translational and rotational dynamics of non-spherical molecules are considerably more complicated due to the involvement of orientational mode.[7]

Hydrodynamics of spheres provides us with simple expressions which have been tested. Hydrodynamics of non-spherical molecules are far more complicated.[8] The situation becomes more complicated when the molecule is not only non-spherical but also a dipole.

In many cases, we come across molecules with distributed charges and ions. None of the closed form theoretical expressions is valid and at this point of time simulations provide the approach to discuss such cases.

Most of the systems of interest are either in water or alcohol which are the universal solvents. Water itself exhibits quite complex rotational and translational dynamics.[9] This mechanism has been studied recently. Non-spherical molecules with distributed charges couple strongly to the jump motions of water. The coupling between water and ions has been studied from nitrate, sulphate and acetate recently.[10]

There is an important class of small linear molecules like carbon monoxide, nitric oxide and cyanide whose motions have not been studied in detail. These molecules are comparable to the size of water. These types of non-spherical molecules can be considered as ellipsoids. The present work focuses on the coupled water-solute dynamics of linear molecules of carbon monoxide, nitric oxide and cyanide in water. The dynamics of these linear molecules display dynamical feature that is expected to be different from those of the spherical ions or molecules like K, Na and Cl. An important point is that the coupling of water to the small molecules can give rise to exotic dynamics which has not been studied. These molecules, in particular, CO and $CN^-$ have been studied spectroscopically in great detail. CO and $CN^-$ fall in the range which could have been studied intensively. However, many of the studies ignored to take a detailed look at the motion of these kinds of ions.

We explore the detailed mechanism of coupled jumps of the solvent and solute and calculate the orientational time correlation function to study the dynamics of orientation. We also examine the validity of hydrodynamic predictions. We develop a mode coupling theory approach and is applied to study the translation-rotation coupling.

## II. SYSTEM AND SIMULATION DETAILS

Molecular dynamics simulations of carbon monoxide (CO), nitric oxide (NO) and cyanide ($CN^-$) in water have been performed using the LAMMPS package.[11] We have considered three separate systems of CO, NO and $CN^-$ in water molecules in a cubic simulation cell of length 51.2 Å (which is equivalent to 0.1 M concentration). We have used extended simple point charge (SPC/E) model of water.[12]

For the simulation of CO in water, 8 CO molecules have been taken with 4444 water molecules in a cubic simulation cell of length 51.2 Å. CO is modelled as a two-site rigid body having partial charges and Lennard-Jones parameters on either site. The force field parameters used for CO are given in **Table I[(a)]**.[13,14] The simulations have been performed under periodic boundary conditions with a cut-off radius of 12 Å. The long-range forces have been computed with the Ewald summation. To obtain the interaction between unlike species, we have used the Lorentz-Berthelot combination rules, i.e. $\varepsilon_{ij} = \sqrt{\varepsilon_i \varepsilon_j}$ and $\sigma_{ij} = \frac{\sigma_i + \sigma_j}{2}$.

The system has been equilibrated for 400ps at constant temperature and pressure (NPT) and later in NVT ensemble for 200ps. In order to apply bond and angle constraints, SHAKE algorithm has been incorporated. We have studied the system at both 300 K and 270 K separately. The final production runs have been carried out in NVT ensemble for 5ns. The equations of motions were integrated using the velocity-Verlet integrator with an MD time step of 1 fs. Coordinates were stored for every 2fs to evaluate various properties.

To simulate NO in water, we have used the same system size and concentration as that of CO in water. The force field parameters used for NO are given in **Table I[(b)]**.[15] The simulation procedure followed for NO was similar to that of CO.

For the cyanide ion, we have considered a two-site rigid body model of 8 CN[-] in 4444 water molecules. The geometric optimization of cyanide was done by density functional theory (DFT) method using 6-311G basis set at the B3LYP level of theory.[16] Since CN[-] in water is charged, we have added a counter-ion K[+] in order to neutralize the system. The self-interaction parameters used are reported in **Table I[(c)]**.[17] We have followed the same simulation procedure of CO for CN[-] system.

**TABLE I.** Force field parameters of CO, NO and CN⁻

| Molecule | Atom(i) | $\sigma_{ii}$(Å) | $\varepsilon_{ii}$(kcal/mol) | $q_i$(e) | Equilibrium bond length(Å) |
|---|---|---|---|---|---|
| (a) CO | $C_{CO}$ | 3.490 | 0.0453 | 0.0203 | |
| | $O_{CO}$ | 3.130 | 0.1261 | -0.0203 | $b_{CO} = 1.128$ |
| (b) NO | $N_{NO}$ | 3.014 | 1.579 | 0.0288 | |
| | $O_{NO}$ | 2.875 | 1.926 | -0.0288 | $b_{NO} = 1.15$ |
| (c) CN⁻ | $C_{CN}$ | 3.852 | 0.105 | -0.553 | |
| | $N_{CN}$ | 3.660 | 0.069 | -0.447 | $b_{CN^-} = 1.172$ |
| | K⁺ | 3.527 | 0.0869 | +1.000 | |

We have also verified CN⁻ system using a 3-point charge model. A distributed multipole analysis (DMA) using GDMA program is employed for the spatial distribution of charge for this model. We have carried out density functional theory (DFT) calculations at the B3LYP/aug-cc-PVQZ level to calculate the multipole moments for CN⁻ [17]. The three site model consists of Lennard-Jones interaction sites and a massless point charge placed at the centre of mass of carbon and nitrogen atoms. We have studied the CN⁻ system with the charges thus obtained, i.e., -0.447e on C and -0.553e on N. However, the structural and diffusive properties of the three-site model of CN⁻ do not show significant variation from the two-site model.

## III. HYDRODYNAMIC PREDICTIONS

In this section we present the theoretical description of the hydrodynamic prediction for both the translational and rotational motion of rod-shaped molecules (prolates).

### A. Translational diffusion

According to the stick hydrodynamic theory, for rod-shaped molecules, the translational diffusion coefficient in the direction parallel to the major axis ($D_\parallel$) is twice to that in the perpendicular direction (D⊥).[18] However, the slip hydrodynamic theory predicts that the ratio of the value of D∥ to D⊥ value approaches the aspect ratio ($\kappa$).[19,20] The expressions for D∥ and D⊥ for a rod-like molecule in a solvent of viscosity η are given by,[18]

$$D_{\parallel} = \frac{k_B T \ln(a/b)}{2\pi\eta a}$$
$$D_{\perp} = \frac{k_B T \ln(a/b)}{4\pi\eta a}$$
(1)

where a and b are the length and diameter of the rod-shaped molecules respectively and the aspect ratio $\kappa = a/b$.

The stick hydrodynamic predictions for translational diffusion for ellipsoids are well known and are given by,[21]

$$D_{\parallel} = \frac{k_B T \left[(2a^2 - b^2)S - 2a\right]}{(a^2 - b^2) 16\pi\eta}$$
(2)

and

$$D_{\perp} = \frac{k_B T \left[(2a^2 - 3b^2)S + 2a\right]}{(a^2 - b^2) 32\pi\eta}$$
(3)

where a and b are the lengths of the semi-major and semi-minor axes of the ellipsoid, respectively and

For translational diffusion, S is defined for a prolate as follows, [21,22]

$$S_P = \frac{2}{(a^2 - b^2)^{1/2}} \log \frac{a + (a^2 - b^2)^{1/2}}{b}$$
(4)

### B. Rotational diffusion

From the stick hydrodynamic theory, the rotational diffusion coefficient for the prolate is given as follows:[23]

$$D_R = \frac{3}{2} \frac{\kappa\left[(2\kappa^2-1)S - \kappa\right]}{\kappa^4 - 1} D_s \qquad (5)$$

where

$$D_s = \frac{k_B T}{6V\eta} \qquad (6)$$

Here V is the volume of the ellipsoid given by $V = \frac{4}{3}\pi a^2 b$.

For rotational diffusion, S is defined for prolate as follows:[24]

$$S_P = (\kappa^2-1)^{-1/2} \ln\left[\kappa + (\kappa^2-1)^{1/2}\right] \qquad (7)$$

The detailed study of rotational relaxation of the non-spherical molecule could not be executed till Hu and Zwanzig obtained the results for the slip boundary conditions. However, no analytical expression exists for the slip boundary conditions. We have used the table provided by the Hu and Zwanzig for the calculation of rotational diffusion under slip boundary conditions.[25]

## IV. MODE-COUPLING THEORY FOR DIATOMIC MOLECULES IN WATER : EFFECTS OF TRANSLATION-ROTATION COUPLING

As mentioned in the introduction, a theoretical calculation of translational and rotational diffusion (or, mobility) of a diatomic ion is hard because not only the molecules are aspherical, but also translational motion is coupled to rotational motion and vice versa. One can however make certain assumptions to simplify the problem. Towards this goal, we note that the charge-dipole and charge-charge interactions vary on a length scale longer than the non-polar Lennard-Jones type interactions that dominate the short range interactions. This separation in the range of forces allows us to separate the total friction (which is approximately given by the integration over force-force time correlation function). We can therefore decompose the total friction into two terms in the following fashion

$$\zeta_{T,ion} = \zeta_{T,Stokes} + \zeta_{T,DF}$$
$$\zeta_{R,ion} = \zeta_{R,DSE} + \zeta_{R,DF}$$

(8)

The translational friction $\zeta_{T,Stokes}$ arises due to shear viscosity of the solvent and obeys Stokes' law. Here, this contribution of Stokes' friction is approximated as $4\pi\eta_0 R_{ion}$, where $\eta_0$ is the viscosity of the solvent and $R_{ion}$ is the radius of the ion. $\zeta_{R,DSE}$ is the same for rotational friction which, under stick boundary condition is $8\pi\eta_0 R_{ion}^3$. The remaining two terms on the right side of **Eq. (8)** are the contributions due to polar interactions. The translational dielectric friction, $\zeta_{T,DF}$, is given by the force-force time correlation function while the rotational dielectric friction, $\zeta_{R,DF}$, is given by the torque-torque time correlation function. It is highly non-trivial to calculate these quantities for non-spherical polar (charged or dipolar) molecules in water. In the case of diatomic molecules, the self-rotational and self-translational motions complicate the descriptions. This problem also provides a nice example of translation-rotation coupling – these are not independent functions.

We now proceed to derive expressions for these frictions.

## A. Translational Friction

We start with the Kirkwood expression for the translational dielectric friction in terms of the ion-solvent force-force correlation function. We first consider the general case.

$$\zeta_{DF} = \frac{1}{3k_BT}\int_0^\infty dt \frac{1}{V}\int d\mathbf{r} \frac{1}{4\pi^2}\int d\Omega \langle \mathbf{F}(r,\Omega,0).\mathbf{F}(r,\Omega,t)\rangle \qquad (9)$$

Classical density functional theory (DFT) offers an expression for the force density on the ion located at position r at time t:[26,27]

$$\mathbf{F}(r,t) = k_B T\, n_{ion}(\mathbf{r},\mathbf{\Omega},t)\nabla\int d\mathbf{r}'d\Omega' c(\mathbf{r},\mathbf{r}',\mathbf{\Omega},\mathbf{\Omega}')\delta\rho(\mathbf{r}',\mathbf{\Omega}',t) \qquad (10)$$

In the case of a spherical ion, both expressions (9) and (10) becomes somewhat simpler because the ion does not need an orientational variable. Let us discuss this simple case first. By using standard Gaussian approximation, microscopic expression of the dielectric friction is obtained as

$$\zeta_{T,DF}(z) = \frac{2k_BT\rho_0}{3(2\pi)^2}\int_0^\infty dt\, e^{-zt}\int_0^\infty dk\, k^4 F_{ion}(k,t)\left|c_{id}^{10}(k)\right|^2 F_{solvent}^{10}(k,t) \qquad (11)$$

This was first derived by Bagchi and co-workers.[26] Here $c_{id}^{10}(k)$ and $F_{solvent}^{10}(k,t)$ are the longitudinal components of the ion-dipole direct correlation function (DCF) and the intermediate scattering function (ISF) of the pure solvent, respectively. $\rho_0$ is the average number density of the solvent. $F_{ion}(k,t)$ denotes the self-intermediate scattering function of the ion which is defined as

$$F_{ion}(k,t) = \frac{1}{N}\langle \rho_{ion}(k,t)\rho_{ion}(k,0)\rangle \qquad (12)$$

where $\rho_k$ is the Fourier component of density. $F_{solvent}^{10}(k,t)$ is obtained from inverse Laplace transform of $F_{solvent}^{10}(k,z)$ which is expressed as

$$F_{solvent}^{10}(k,z) = \frac{S_{solvent}^{10}(k)}{z + \Sigma_{10}(k,z)} \tag{13}$$

where $\Sigma_{10}$ is the generalized rate of relaxation of dynamic structure factor, given by

$$\Sigma_{10}(k,z) = \frac{2k_B T f(110;k)}{I[z + \Gamma_R(k,z)]} + \frac{k_B T k^2 f(110;k)}{m\sigma^2 [z + \Gamma_T(k,z)]} \tag{14}$$

where $\Gamma_R(k,z)$ and $\Gamma_T(k,z)$ are the wavenumber and frequency dependent rotational and translational memory kernels, respectively. The term, $f(110;k)$ appeared in **Eq. (14)** is defined as the orientational correlation function that is related to the direct correlation function, $c(110,k)$.

$$f(110;k) = 1 - \frac{\rho_0}{4\pi} c(110;k) \tag{15}$$

The contribution of $f(110;k)$ in the single particle limit ($k \to \infty$) become unity ($f(110;k)=1$), but it has a significant role in the collective density relaxation.

The self-intermediate scattering function of the ion, $F_{ion}(k,t)$ is obtained from inverse Laplace transform of $F_{ion}(k,z)$. In the case of a spherical solute, rate of relaxation of dynamic structure factor ($\Sigma_{10}$) have contribution only from translation motion

$$F_{ion}(k,z) = \frac{1}{z + D_{ion}^T k^2} \tag{16}$$

where $D_{ion}^T$ is again determined from the total friction by Stokes-Einstein relation. Therefore, **Eq. (11)** being a mode-coupling, nonlinear equation has $\zeta(z)$ on both sides that has to be solved self-consistently.

The above equations neglect the rotational self-motion of the tagged ion whose friction is being studied. This requires inclusion of the orientation $\Omega$ in **Eqs. (9)** and **(10)**. The resulting expressions are complex, with **Eq. (11)** as the first term of a series of terms that originate from spherical expansion of two densities and the direct correlation function, and integration over all the variables.[28] While numerical studies have indeed been carried out with **Eq. (11)**, no study of the detailed equations. However, these equations serve to explain at a molecular level how rotational jump can lower translational friction. Expression for the frequency dependent friction is given by

$$\zeta(z) = \zeta_{bare} + A\int_0^\infty dt\, e^{-zt} \int_0^\infty dk\, k^2 \sum_{l_1 l_2 m} F_{l_1 m}^s(k,t) c_{l_1 l_2 m}^2(k) F_{l_2 m}(k,t) \qquad (17)$$

where A is a numerical constant. For spherical ion, this expression reduces to **Eq. (11)** given above. The above expression helps to understand the role of orientational self-motion in lowering the dielectric friction.

## B. Rotational Friction:

One can derive microscopic expressions for dielectric friction by using the same procedure as detailed above, except that we need to deal with torque whose evaluation is more involved. The rotational friction experienced by the diatomic solute can be expressed as a time integral of torque-torque time correlation function (TTTCF)

$$\zeta_{R,DF} = \frac{1}{2k_B T} \int_0^\infty dt\, \frac{1}{V}\int d\mathbf{r}\, \frac{1}{4\pi^2}\int d\Omega \langle \mathbf{N}(r,\Omega,0).\mathbf{N}(r,\Omega,t)\rangle \qquad (18)$$

where N(Ω,t) is the total torque experienced by the solute molecule with orientation Ω at time t.

Similar to the expression of force (**Eq. (10)**), classical density functional theory provides an elegant expression of torque on the point dipole due to density fluctuation

$$\mathbf{N}(r,\Omega,t) = k_B T\, n_{ion}(\mathbf{r},\mathbf{\Omega},t) \nabla_\Omega \int d\mathbf{r}' \int d\mathbf{\Omega}'\, c(\mathbf{r}-\mathbf{r}',\mathbf{\Omega},\mathbf{\Omega}')\delta\rho(\mathbf{r}',\mathbf{\Omega}'t) \tag{19}$$

where $\delta\rho(\mathbf{r},t) = \rho(\mathbf{r},t) - \rho_0$ is the density fluctuation around the average density $\rho_0$. $c(\mathbf{r}-\mathbf{r}',\Omega)$ is the direct correlation function between the point dipole at (r,Ω) and the solvent density at $r'$. The direct correlation function can be expanded in spherical harmonics to obtain

$$N(r,\Omega,t) = \frac{k_B T}{(2\pi)^3} \sum_{l_1 m_1} A_{l_1 m_1}(r,t) Y_{l_1 m_1}(\Omega) \sum_{l_2 m_2} \langle \nabla_\Omega Y_{l_2 m_2}(\Omega) \rangle \int d\mathbf{k}\, e^{i\mathbf{k}\cdot\mathbf{r}}\, c_{l_2 l_3 m_3}(\mathbf{k})\, \delta\rho_{l_3 m_3}(\mathbf{k},t) \tag{20}$$

Substituting **Eq. (20)** in **Eq. (18)** gives the final expression for the memory functions of the rotational friction

$$\Gamma_c(z) = \Gamma_{bare} + A \int_0^\infty dt\, e^{-zt} \int_0^\infty dk\, k^2 \sum_{l_1 l_2 m} F^s_{l_1 m}(k,t) c^2_{l_1 l_2 m}(k) F_{l_2 m}(k,t) \tag{21}$$

where the constant $A = \frac{\rho}{2(2\pi)^2}$. $c_{l_1 l_2 m}(k)$ is the $l_1 l_2 m^{th}$ coefficient of the direct correlation function between solute and solvent and $F^s_{l_1 m}(k,t)$ and $F_{l_2 m}(k,t)$ are self and cross terms of orientational correlation function, defined as

$$F^s_{l_1 m}(k,t) = \langle e^{i\mathbf{k}\cdot(\mathbf{r}_i(t)-\mathbf{r}_i(0))} Y_{lm}(\Omega_i(0)) Y_{lm}(\Omega_i(t)) \rangle \tag{22}$$

$$F_{l_2 m}(k,t) = \sum \langle e^{i\mathbf{k}\cdot(\mathbf{r}_j(t)-\mathbf{r}_i(0))} Y_{lm}(\Omega_i(0)) Y_{lm}(\Omega_j(t)) \rangle \tag{23}$$

The **Eqs. (20) - (22)** show that the rotational friction is coupled to translational diffusion because the memory functions contain both rotational and translational friction terms. In addition, these equations are to be solved self-consistently (boot-strapping) that further couples rotational and translational diffusion coefficients.

Earlier such calculations were performed for spherical molecules with point dipoles and also spherical ions in a solvent of spherical dipolar molecules.[29] However, detailed calculations with linear diatomics with distributed charges are yet to be carried out. However, the above equations bring out the essence of the observed translational-rotational coupled motions that are observed in computer simulations discussed below.

## V. RESULTS AND DISCUSSION

### A. Carbon Monoxide

#### 1. Structure

Radial distribution function

In order to analyse the structure of the system, we have studied the radial distribution functions(RDFs). **Figure 1** shows the RDF of oxygen atoms of CO ($O_{CO}$) with oxygen atoms of water ($O_w$) and hydrogen atoms of water ($H_w$) at 300K. The radial distribution function of $O_w$ and $H_w$ around carbon atoms of CO ($C_{CO}$) is also shown. The RDF shows orientational anisotropic solvation by water. The first maximum of the $C_{CO} - O_W$ RDF is at 3.6 Å. It is observed that the number density of $O_w$ around $O_{CO}$ and $C_{CO}$ is more than the distribution of $H_w$ around $O_{CO}$ and $C_{CO}$. So the weak correlation of $H_W$ with $O_{CO}$ and $C_{CO}$ shows that the probability of finding $H_W$ around $O_{CO}$ and $C_{CO}$ is less than the probability of $O_{CO}$ and $C_{CO}$ being surrounded by $O_W$. It is observed from the RDFs at 300K and 270K(not shown here) that the temperature dependence of the structural quantity is quite weak. But it is observed that in the RDF of $O_{CO}$ and $O_W$, the value of the first peak and second peak increases slightly with the decrease in temperature. This shows that the correlation of CO with water molecules increases with the drop in temperature.

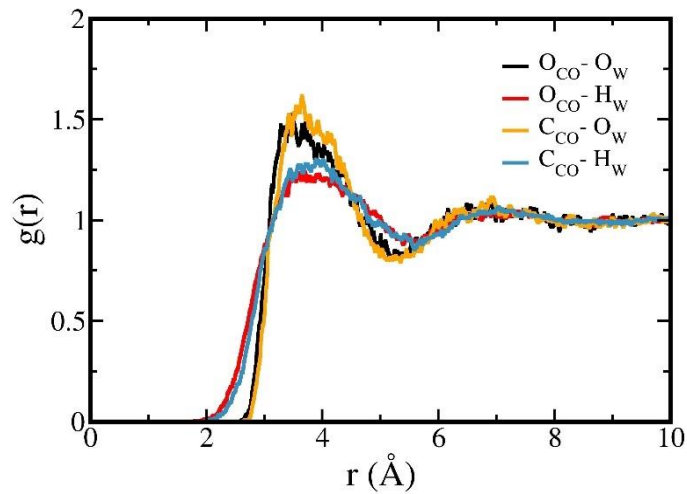

**FIG 1. Radial distribution functions (RDFs) of water oxygen ($O_W$) and water hydrogen ($H_W$) atoms around oxygen($O_{CO}$) and carbon($C_{CO}$) atoms of carbon monoxide at 300K**

### 2. Diffusion

In order to study the translational dynamics, we have computed the mean-square displacements of the center of mass of both CO and water at 300K and 270K. The self-diffusion constant, D can be determined from mean-square displacement by the use of Einstein's diffusion equation,

$$D = \lim_{t \to \infty} \frac{1}{6t} \langle |r(t) - r(0)|^2 \rangle \qquad (24)$$

where r(t) is the position of the center of mass at time t.

**Table II** shows the simulated value of self-diffusion constants of CO and water at 300K and 270K. As expected the diffusion coefficient increases with the increase in temperature. The diffusion constants at both the temperatures clearly show the coupling of CO to water. The diffusion coefficient of CO at both the temperatures agrees satisfactorily with the observed experimental diffusion constants, i.e. 2.43 x 10$^{-5}$ cm$^2$ s$^{-1}$ (at 303 K) and 1.07 x 10$^{-5}$ cm$^2$ s$^{-1}$ (at 283K).[30]

**TABLE II. Self-diffusion constants of CO and water at 300K and 270K**

| Temperature (K) | Diffusion coefficient (1 x 10$^{-5}$ cm$^2$ s$^{-1}$) | | |
|---|---|---|---|
| | CO (Experimental) | CO (our work) | SPC/E water |
| 300 | 2.43 (303K) | 2.68±0.07 | 2.65±0.01 |
| 270 | 1.07 (283K) | 1.26±0.02 | 1.29±0.02 |

Further information of the dynamics can be provided by the self-part of the van Hove correlation function $G_s(r,t)$:[31,32]

$$G_s(r,t) = \frac{1}{N}\left\langle \sum_{i=1}^{N} \delta\left[r - |r_i(t) - r_i(0)|\right] \right\rangle \quad (25)$$

where $r_i(t)$ is the position of the i$^{th}$ particle at time t. The self-correlation function $G_s(r,t)$ provides a description for the motion of the individual particles in detail. The r dependence of $G_s(r,t)$ is shown in **figure 2(a)** and the r dependence of $r^2 G_s(r,t)$ is shown in **figure 2(b)**.

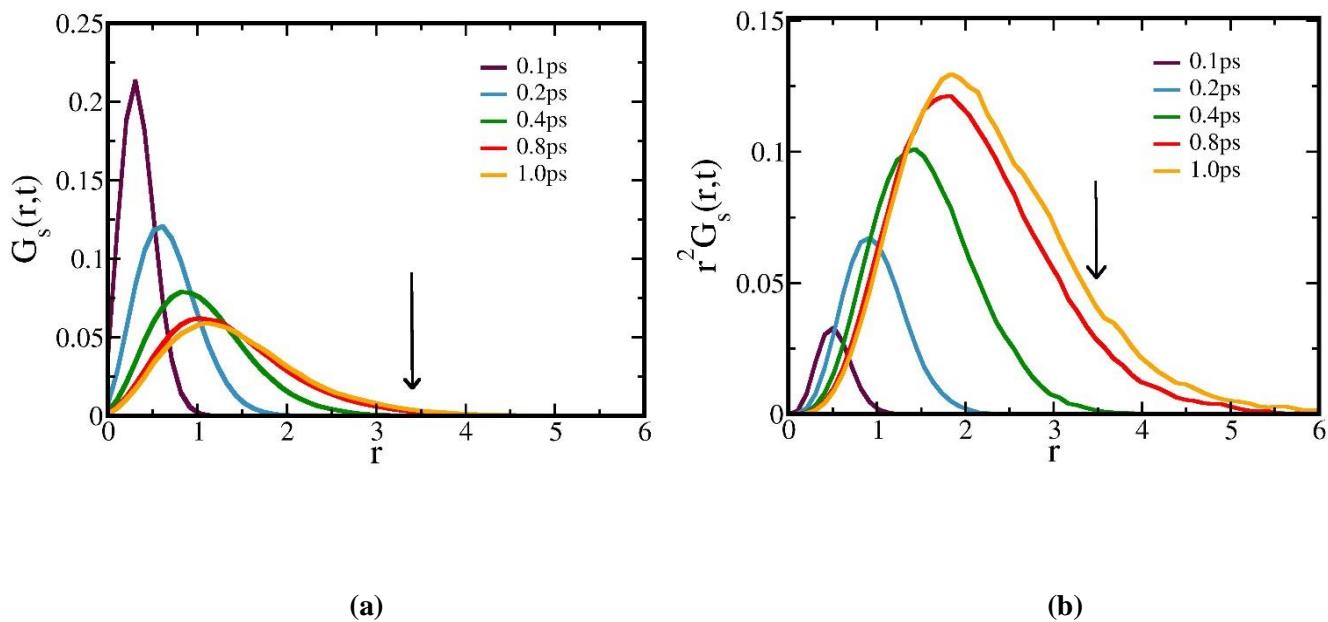

(a)                          (b)

**FIG 2. Self –part of the van-Hove correlation function $G_s(r,t)$ as a function of r for t=0.1, 0.2, 0.4, 0.8 and 1.0 ps**

We find a significant deviation from the Gaussian nature. Any emergence of a secondary peak is considered as an occurrence of the hopping process where the particle jumps to the formerly occupied positions by one of its nearest neighbors. Nonetheless, the absence of such a secondary peak does not allow us to conclude that there is no contribution from the hopping processes. It has been shown in many previous studies that the non-Gaussian nature of the probability distribution of displacement suggests the dynamic heterogeneity of the system.

We have also examined the single CO molecule trajectory and the trajectory of the nearest neighboring water of the corresponding CO molecule.[33] **Figure 3** shows few trajectories of square displacement of two CO molecules and its neighboring water from the initial time of trajectory. The trajectories show slaving of the motion of small solutes to the dynamics of water. A correlated jump motion of CO is observed in these trajectories.

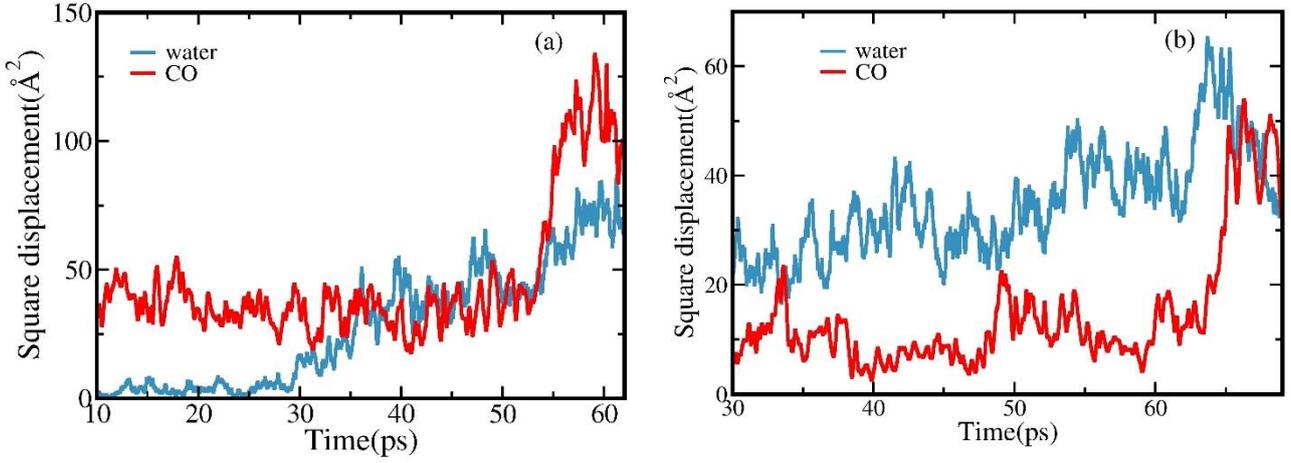

**FIG 3. Square displacement of two different CO molecules and the nearest neighboring water molecule. The figure shows a jump motion in CO molecule and its neighboring water molecule.**

### 3. Rotational dynamics

In order to examine the dynamics of orientation, we have analysed the orientational time correlation function. The orientational time correlation function of CO at both 300K and 270K has been studied. The rotational relaxation time can be measured by computing the first-rank and second-rank orientational time correlation function, $C_l(t)$,[34] which is defined as

$$C_l(t) = \left\langle P_l(\vec{r}(t_0).\vec{r}(t_0+t)) \right\rangle \tag{26}$$

where $P_l(x)$ is the $l^{th}$ Legendre polynomial defined as

$$\begin{aligned}P_1(x) &= x \\ P_2(x) &= \frac{1}{2}(3x^2 - 1)\end{aligned} \tag{27}$$

Here $\vec{r}$ is a unit vector along the bond vector. Angular brackets stand for ensemble averaging. The $l^{th}$ rank orientational correlation time $\tau_l$ is defined as

$$\tau_l = \int_0^\infty C_l(t)dt \qquad (28)$$

Where $C_l(t)$ is the $l^{th}$ rank orientational time correlation function.

## B. Nitric Oxide

### 1. Structure

**Figure 4** shows the RDF of oxygen atoms ($O_{NO}$) and nitrogen atoms ($N_{NO}$) of NO with oxygen atoms of water ($O_W$) and hydrogen atoms of water ($H_W$) at 300K.

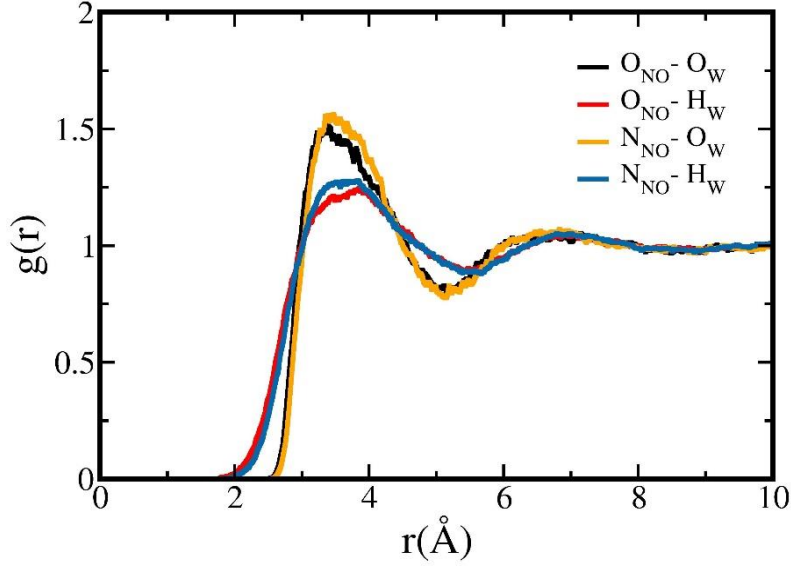

**FIG 4. Radial distribution functions of water oxygen ($O_W$) and water hydrogen ($H_W$) atoms around oxygen ($O_{NO}$) and nitrogen ($N_{NO}$) atoms of nitric oxide at 300K.**

### 2. Diffusion

**Table III** shows the simulated value of self-diffusion constants of NO and water at 300K and 270K. The diffusion coefficient values of NO in water agree satisfactorily with the observed experimental results. At 300K, we obtained a diffusion coefficient value of $(3.48 \pm 0.10) \times 10^{-5}$

cm$^2$ s$^{-1}$ compared to the experimental value 3.96 x 10$^{-5}$ cm$^2$ s$^{-1}$(at 303K). In the NO-water system, NO being a nearly neutral molecule which is hydrophobic in nature, enhances the mobility of the NO molecule in the mixture. Thus the diffusion coefficient of NO is higher than water in the NO-water mixture.

**TABLE III. Self-diffusion constants of NO and water at 300K and 270K**

| Temperature (K) | Diffusion coefficient (1 x 10$^{-5}$ cm$^2$ s$^{-1}$) | | |
|---|---|---|---|
| | NO (Experimental)[30] | NO (our work) | SPC/E water |
| 300 | 3.96 (303K) | 3.48 ± 0.10 | 2.70 ±0.03 |
| 270 | 1.55 (283K) | 1.53±0.02 | 1.29± 0.01 |

**Figure 5** shows few trajectories of square displacement of two NO molecules with its neighboring water from the initial time of trajectory. It is evident from the figure that a translational jump takes place in NO which enhances its translational motion.

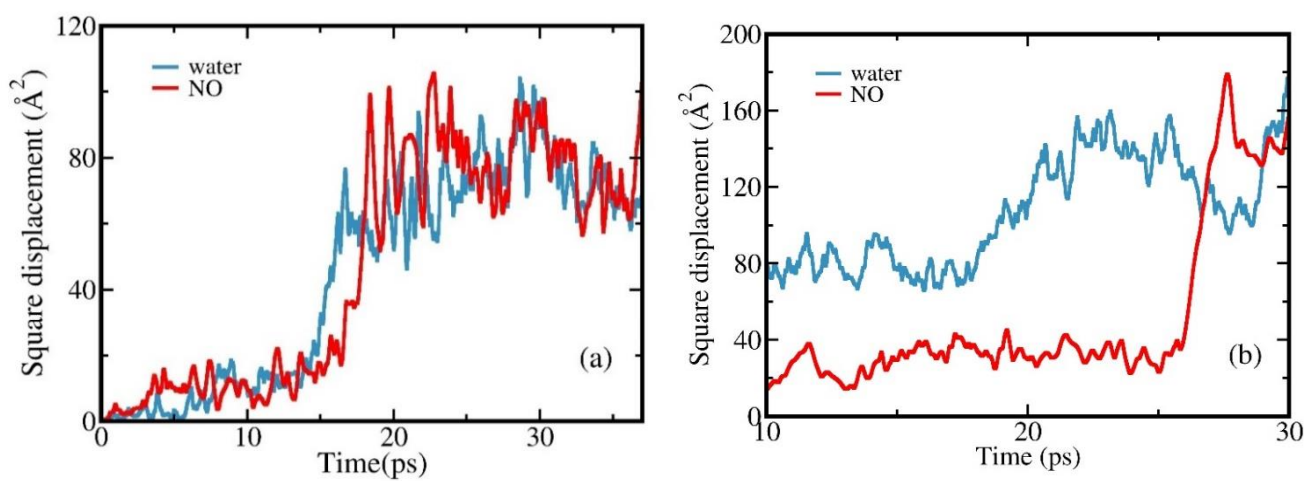

**FIG 5. Square displacement of two different NO and the nearest neighboring water molecule. The figure shows a jump motion in NO molecule and its neighboring water molecule.**

## C. Cyanide

### 1. Structure

**Figure 6** shows radial distribution functions of carbon atoms(C) and nitrogen atoms(N) of CN$^-$ with oxygen atoms of water (O) and hydrogen atoms of water (H) at 300K. The RDF shows the rotational isotropy in CN$^-$.[35]

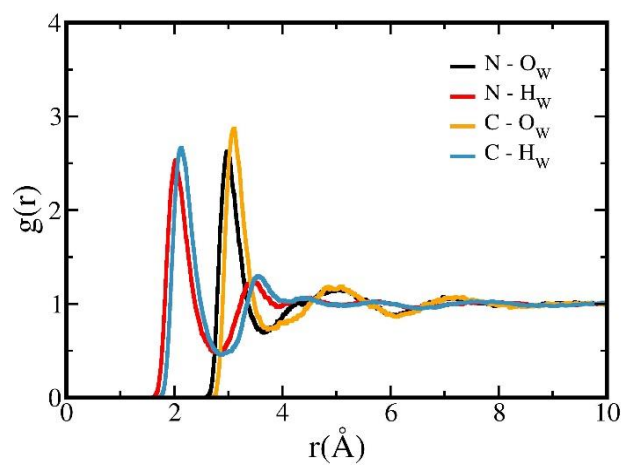

**FIG 6. Radial distribution functions of oxygen atoms of water ($O_W$) and hydrogen atoms of water ($H_W$) atoms around nitrogen(N) and carbon(C) atoms of cyanide at 300K.**

### 2. Diffusion

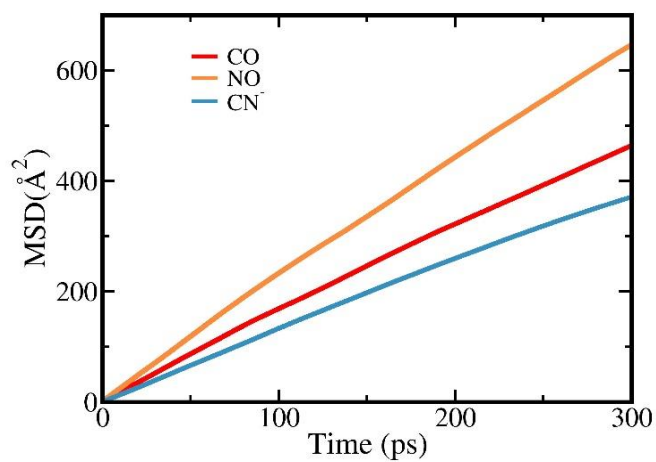

**FIG 7. Plot of MSD versus time of CO, NO and CN$^-$ at 300K**

**Table IV** shows the simulated value of self-diffusion constants of CN$^-$ and water at both the temperatures. As the water around CN$^-$ is more structured, the diffusion of CN$^-$ becomes slow. The diffusion coefficient of CN$^-$ at 300K agrees satisfactorily with the observed experimental diffusion constant, i.e. 2.077 x 10$^{-5}$ cm$^2$ s$^{-1}$ (infinite dilution at 298K).[36]

**TABLE IV. Self-diffusion constants of CN$^-$ and water at 300K and 270K**

| Temperature (K) | Diffusion coefficient (1 x 10$^{-5}$ cm$^2$ s$^{-1}$) | | |
|---|---|---|---|
| | CN$^-$ (Experimental) | CN$^-$ (our work) | SPC/E water |
| 300 | 2.077 (298K) | 2.20 ± 0.10 | 2.70 ± 0.03 |
| 270 | - | 0.96 ± 0.04 | 1.29 ± 0.03 |

We have also examined the single solute molecule trajectory and the trajectory of the nearest neighboring water of the corresponding solute molecule. Single CN$^-$ molecule trajectory and the trajectory of the nearest neighboring water of the corresponding CN$^-$ molecule was examined. **Figure 8** shows few trajectories of square displacement of various CN$^-$ molecules and its neighboring water from the initial time of trajectory. The trajectories show slaving of the motion of small solutes to the dynamics of water. A correlated jump motion of CN$^-$ is observed in these trajectories.

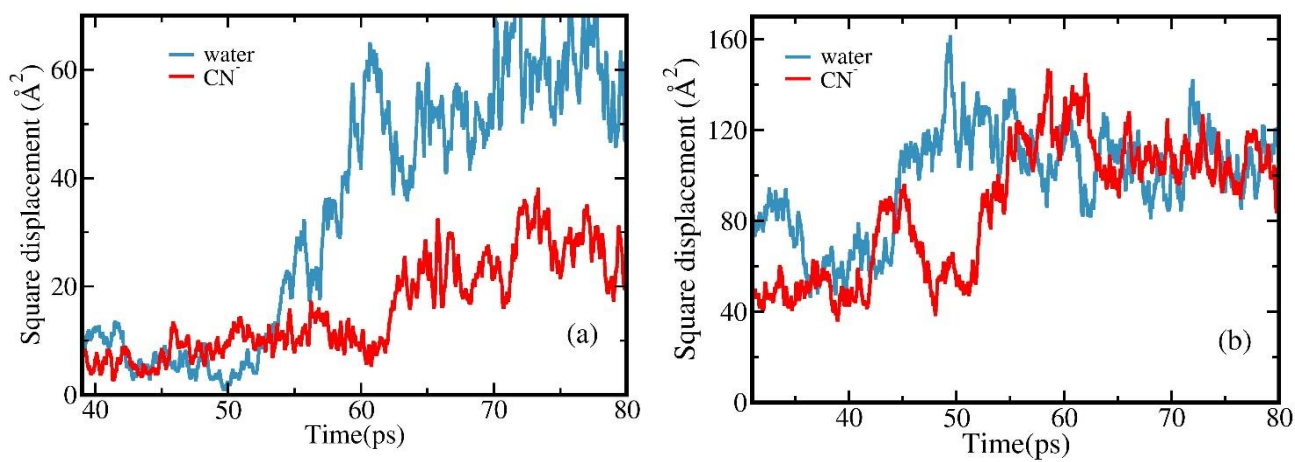

**FIG 8. Square displacement of two different CN⁻ and the nearest neighboring water molecule. The figure shows a jump motion in CN⁻ molecule and its neighboring water molecule.**

### 3. Rotational dynamics

Many previous studies have been done in detail for the relaxation of orientational time correlation function of water. **Figures 9(a)** and **9(b)** show the first-rank and second-rank rotational time correlation function of all the three molecules in water respectively. The first-rank correlation function relaxes much slower than the second-rank correlation function. Additionally, compared to the relaxation of CO and NO, the relaxation of the CN⁻ is the slowest due to the well-structured solvation shell. The faster decay of CO and NO than CN⁻ demonstrates that the rotational motion of the CO and NO in water is faster than the CN⁻ in water.

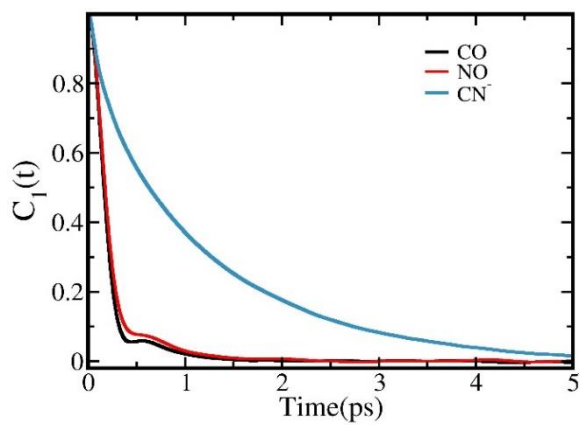

(a)

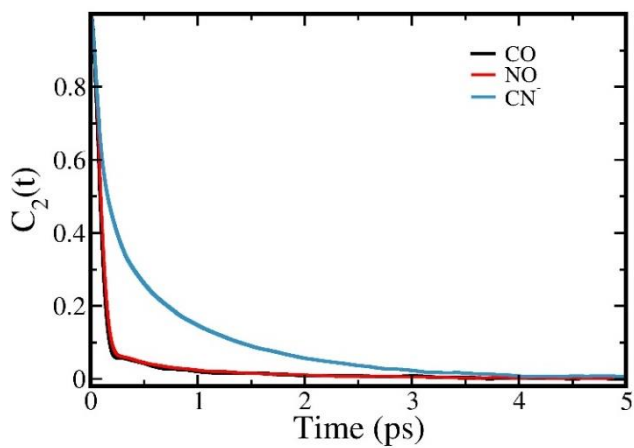

(b)

**FIG 9. (a) First order orientational time correlation function of CO, NO and CN⁻ molecules in water at 300K (b) Second order orientational time correlation function of CO, NO and CN⁻ molecules in water at 300K. The first-rank correlation function relaxes much slower than the second-rank correlation function. It is noted that the relaxation of the CN⁻ is the slowest compared to the relaxation of CO and NO molecules.**

The Debye model of rotational diffusion predicts the $[l(l+1)]^{-1}$ dependence for $\tau_l$, and explicitly the ratio of the first rank orientational correlation time to the second-rank orientational time, $\tau_{l=1}/\tau_{l=2}$, to be equal to three. **Table V** shows the two decay constants $\tau_1$ and $\tau_2$ of CO, NO and CN⁻ at 300K and 270K obtained from the first-rank and second-rank correlation functions. The ratio $\tau_1/\tau_2$ for the three molecules are also listed in **Table V**. A significant deviation from the ratio value is considered to be a signature of the breakdown of the Debye model.[37]

**TABLE V: Decay constants $\tau_1$ and $\tau_2$ and ratio $\tau_1/\tau_2$ of first-rank orientational time correlation function $C_1(t)$ and second-rank orientational time correlation function $C_2(t)$**

| Molecule | Temperature (K) | $\tau_1$ | $\tau_2$ | $\tau_1/\tau_2$ |
|---|---|---|---|---|
| CO | 300 | 0.2116 | 0.1494 | 1.4160 |
|    | 270 | 0.2478 | 0.1834 | 1.3510 |
| NO | 300 | 0.2458 | 0.1730 | 1.4208 |
|    | 270 | 0.2835 | 0.2302 | 1.2315 |
| CN⁻ | 300 | 1.1480 | 0.5641 | 2.0351 |
|     | 270 | 1.6379 | 1.0103 | 1.6212 |

### D. Rotational diffusivity

The rotational motion of linear molecules can be computed from angular velocity autocorrelation function which is defined as

$$C_\omega(t) = \langle \omega(0).\omega(t) \rangle \quad (29)$$

where $\omega(t)$ is the angular velocity vector at time t. By integrating the angular velocity correlation function, the rotational diffusion coefficient can be calculated as [38]

$$D_R = \frac{k_B T}{I} \int_0^\infty \frac{C_\omega(t)}{C_\omega(t=0)} dt \quad (30)$$

where I is the mass moment of inertia of the molecule which is determined by adding the moment of each atom within the molecule about the centre of mass.

We have calculated the rotational diffusion coefficient of CO, NO and CN⁻ molecules using **Eqs. (29)** and **(30)**. We have also compared this value with the rotational diffusion value obtained from the second rank orientational time correlation function. The comparison of the rotational diffusivities obtained from Green-Kubo formula and also from the orientational correlation time is listed in **Table VI**. This shows the rotational motion of CO and NO in water to be faster than that of CN⁻ in water, which is in good agreement with the plots of first and second order orientational time correlation function, as shown in **figure 9(a)** and **9(b)**.

**TABLE VI. Rotational diffusion coefficient of CO, NO and CN⁻ from Green-Kubo formula and from orientational correlation time.**

| Molecule | $D_R = \dfrac{k_B T}{I} \int_0^\infty \dfrac{C_\omega(t)}{C_\omega(t=0)} dt$ (ps⁻¹) | $D_R = \dfrac{1}{6\tau_2}$ (ps⁻¹) |
| --- | --- | --- |
| CO | 1.119 | 1.115 |
| NO | 0.836 | 0.963 |
| CN⁻ | 0.065 | 0.295 |

## E. Translational and rotational diffusion from hydrodynamic predictions

In this section we present the values of $D_\parallel$, $D_\perp$ and $D_{total}$ from the hydrodynamic predictions for prolate-like system. We estimate the dimensions of the carbon monoxide using the known bond length, 1.128 Å, and the radii of carbon and oxygen. In a similar way, using the bond length of NO, 1.15 Å and of CN⁻, 1.172 Å and the corresponding radii [**Table I**], we consider CO, NO and CN⁻ in water systems to be like prolate spheroids.

**Table VII** shows the values of $D_\parallel$, $D_\perp$ and $D_{total}$ for CO, NO and CN⁻ molecules obtained from slip and stick hydrodynamic boundary conditions and also diffusion coefficient from our simulation. The values of total translational diffusion coefficient obtained from slip and stick boundary conditions are in good agreement with our simulation results.

**TABLE VII.** We show the values of $D_\parallel$, $D_\perp$ and $D_{total}$ for CO, NO and CN⁻ molecules obtained from slip and stick hydrodynamic boundary conditions and from our simulation results. It is noted that for CO, NO and CN⁻ in water systems, the values of total translational diffusion coefficient obtained from slip and stick boundary conditions are in good agreement with our simulation results.

| Molecule | SLIP | | | STICK | | | SIMULATION |
|---|---|---|---|---|---|---|---|
| | $D_\parallel$ (x $10^{-5}$ cm²/s) | $D_\perp$ (x $10^{-5}$ cm²/s) | $D_{Tot}$ (x$10^{-5}$ cm²/s) | $D_\parallel$ (x$10^{-5}$ cm²/s) | $D_\perp$ (x$10^{-5}$ cm²/s) | $D_{Tot}$ (x$10^{-5}$ cm²/s) | $D_{Simu}$ (x$10^{-5}$ cm²/s) |
| CO | 2.133 | 1.067 | 5.43 | 2.961 | 6.299 | 1.55 | 2.68 |
| NO | 2.107 | 1.054 | 4.86 | 2.621 | 7.00 | 1.66 | 3.48 |
| CN⁻ | 2.036 | 1.018 | 5.38 | 2.902 | 5.80 | 1.45 | 2.20 |

Here, we have also calculated second rank orientational correlation time from hydrodynamics. The rotational diffusion coefficient under stick boundary conditions for the prolate system was calculated using **Eq. (5)**. We have used the table provided by the Hu and Zwanzig for the calculation of rotational diffusion under slip boundary conditions. **Table VIII** shows the $\tau_2$ values of CO, NO and CN⁻ molecules obtained from both stick and slip hydrodynamic conditions and from our simulation results.

**TABLE VIII.** We show the values of $\tau_2$ for CO, NO and CN⁻ molecules obtained from slip and stick hydrodynamic boundary conditions and from our simulation results.

| Molecule | $\tau_2$ | | |
|---|---|---|---|
| | SLIP (ps) | STICK (ps) | SIMULATION (ps) |
| CO | 15.804 | 33.875 | 0.149 |
| NO | 8.092 | 21.083 | 0.173 |
| CN⁻ | 15.290 | 30.665 | 0.564 |

## VI. CONCLUSIONS

In this work, we have presented a study of the coupled translational and rotational dynamics of three linear, small, molecules namely CO, NO and CN⁻ in water. These three diatomic molecules consist of three atoms, C, N and O. All the three molecules have distributed charges, and one is an ion. Together they allow us to capture and differentiate different coupling scenarios.

All the three systems have been studied at two different temperatures. The equilibrium structural properties of CO, NO and CN⁻ in water have been studied from the radial distribution functions. We have also studied the diffusion phenomena and the temperature dependence of the above mentioned three systems. We have observed that the translational motions of CO, NO and CN⁻ are strongly coupled to their rotational dynamics which in turn are coupled to neighboring water molecules. We have also examined the validity of hydrodynamic predictions by modelling these solutes as prolate ellipsoids. The latter is certainly an approximation which, we believe, is not too serious. The resulting outcome is quite surprising. For the translational diffusion, the predicted values of diffusion are found to lie between the stick and slip boundary value predictions. This is expected. However, the predictions for rotational diffusion was found to be vastly different from the simulation results, in so much so that the simulations results are more than one order of magnitude. That is, the rotational correlation time is one order of magnitude smaller than the hydrodynamic predictions.

The reason for such a complete breakdown of hydrodynamic predictions is not totally surprising because hydrodynamics predictions used here ignore the inertial motion completely. And the rotational dynamics of these small linear molecules contain a large component of inertial motion.

The above lacuna can be overcome by using a mode coupling theory approach because the inertial components are included in the memory function. Such a theory has been developed here and applied to study translation-rotation coupling, although detailed numerical implementation of the complex equations has been left for the future.

**ACKNOWLEDGMENTS**

We thank Department of Science and Technology(DST) for partial financial support of this work. B. B. thanks Sir J. C. Bose fellowship for partial support.